\documentclass[revtex]{emulateapj}

\usepackage{graphicx}

\newcommand{\radm}{\rm rad\ m^{-2}}
\newcommand{\kms}{\rm km\ s^{-1}}
\newcommand{\alfven}{Alfv\'en\ }

\shorttitle{Magnetic field in an IV filament}
\shortauthors{Stil \& Hryhoriw}

\begin{document}

\title{Magnetic field strength in an intermediate-velocity ionized filament\\ in the First Galactic Quadrant}

\author{J. M. Stil}
\author{A. Hryhoriw} 
\affil{Department of Physics and Astronomy, University of Calgary}

\begin{abstract}
We investigate the magnetic field in an intermediate-velocity filament for which the H$\alpha$ intensity in the WHAM survey correlates with excess Faraday rotation of extragalactic radio sources over the length of the filament from $b \approx 20\degr$ to $b \approx 55\degr$. The density-weighted mean magnetic field is $2.8\ \pm 0.8\ \mu$G, derived from rotation measures and an empirical relation between H$\alpha$ emission measure and dispersion measure from Berkhuijsen et al. (2006). In view of the uncertainties in the derived magnetic field strength, we propose an alternative use of the available data, rotation measure and emission measure, to derive a lower limit to the \alfven speed, weighted by electron density $n_e^{3/2}$. We find lower limits to the \alfven speed that are comparable to, or larger than the sound speed in a $10^4\ \rm K$ plasma, and conclude that the magnetic field is dynamically important. We discuss the role of intermediate-velocity gas as a locus of Faraday rotation in the interstellar medium, and propose this lower limit to the \alfven speed may also be applicable to Faraday rotation by galaxy clusters.
\end{abstract}


\section{Introduction}

Structure in the interstellar medium (ISM) is in part created by ejection of matter from stars in the form of stellar winds and supernova explosions that transfer significant amounts of matter, energy and momentum to the surrounding gas. This stellar feedback accelerates, compresses and heats the surrounding gas on scales up to hundreds of pc \citep{mckee1977}. This turbulent energy is an essential part of the dynamo mechanism that forms large-scale magnetic fields in galaxies \citep[e.g.][]{kulsrud1999}. In return, a sufficiently strong magnetic field can restrict the dynamics of the ISM through its coupling to charged particles, even if the ionization fraction is low. The ratio of gas pressure to magnetic pressure, or the ratio of the sound speed to the \alfven speed, is used to quantify the dynamical importance of the magnetic field. 

The potential importance of magnetic fields for the topology of the ISM on small scales was recognized more than 50 years ago \citep{shajn1958}, although it has become clear that the relation between filamentary structure and magnetic fields can be complex \citep[][for a review]{crutcher2012}. A lot of work has focused on the effects of expanding super bubbles on scales of hundreds of parsecs \citep[e.g.][]{mckee1977,ferriere1991,tomisaka1998,stil2009,vanmarle2015}. In recent years, parsec-scale filamentary structure was revealed in the local ISM, associated with thin filaments of atomic hydrogen gas that are aligned with the local magnetic field revealed by polarization of starlight \citep{mcclure2006,clark2014}, and also in filaments of interstellar dust \citep{ade2014}. 

Faraday rotation of the plane of polarization of linearly polarized radio waves has proven to be an effective probe of the Milky Way's global magnetic field, as well as small-scale structure down to parsec scales \citep[][for a recent review]{haverkorn2015}. The polarization angle $\theta$ rotates by an amount $\Delta \theta$ that is proportional to the wavelength $\lambda$ of the radio wave, according to the relation
\begin{equation}
\Delta \theta = RM \lambda^2,
\label{RM-eq}
\end{equation}
where $RM$ is the rotation measure derived from observations of linear polarization at multiple wavelengths. For a compact polarized source observed through a foreground screen, $RM$ is independent of wavelength and equal to the Faraday depth $\phi$ between the source and the observer, defined as
\begin{equation}
\phi = 0.81 \int n_e \vec{B}\cdot d \vec{l},
\label{phi-eq}
\end{equation}
with $\phi$ in $\radm$, electron density $n_e$ in $\rm cm^{-3}$, magnetic field strength $B$ in $\rm \mu G$, and path length $l$ in pc. The integral is evaluated along the entire path of the radiation from its source to the observer. Its sign depends on the (mean) direction of the magnetic field projected on the line of sight, positive values indicating that the magnetic field is directed towards the observer. 

The distinction between the observed quantity $RM$ and Faraday depth $\phi$ is not always made explicit. Faraday depth is a quantity that can be different for separate emission regions along the line of sight, or for regions that are not resolved by the beam of the telescope. If there is only a single Faraday depth in the problem, $RM$ is a direct measure of Faraday Depth. In the general case of more than one Faraday depth, the vector superposition of the polarization of the different regions can only be approximated by Equation~\ref{RM-eq} as a first-order Taylor expansion over a short wavelength range. In this case it is possible that $RM$ defined in Equation~\ref{RM-eq} is a function of wavelength, and different from the Faraday depth for any synchrotron emitting region along the line of sight. Although broad-band observations may reveal this effect, it is a source of error when equating Faraday depth to $RM$ measured with a modest bandwidth \citep{farnsworth2011}. Substituting $RM = \phi$ is a good assumption in the case of emission of a compact polarized source with negligible internal Faraday rotation, that travels through a foreground Faraday screen. In this paper, the distinction between $RM$ and $\phi$ is made mainly to emphasize how this assumption enters the derivation presented in Section~\ref{vA-sec}.

Inversion of Equation~\ref{phi-eq}  with $RM = \phi$ to derive the magnetic field from observations is not straightforward.
The mean magnetic field along the line of sight may be estimated by dividing $\phi$ by the dispersion measure $DM$, defined as
\begin{equation}
DM = \int n_e dl,
\label{DM-eq}
\end{equation}
with units $\rm pc\ cm^{-3}$. $DM$ can be measured directly with pulsars. For studies of the global magnetic field of the Milky Way, dispersion measures provide a 3-dimensional model for the electron density \citep{cordes2002} that can be used to derive a dispersion measure in the direction of a radio source for which $RM$ has been measured. For smaller structures in the ISM there is usually not enough information to derive $DM$ directly. In this case, one can attempt to use emission measure,
\begin{equation}
EM = \int n_e^2 dl,
\label{EM-eq}
\end{equation}
as a proxy and use an empirical relation between $EM$ and $DM$ derived for pulsars \citep{berkhuijsen2006}. This approach is restricted to situations where the Faraday rotating plasma is directly observable. 

For a turbulent medium, or in situations when $n_e$ is correlated with the magnetic field, this estimate can provide magnetic field estimates that differ significantly from the actual magnetic field strength \citep{beck2003}. Simulations of a supernova-driven ISM by \citet{deavillez2005} do not show a strong global correlation between magnetic field strength and density. This does not exclude localized conditions in which the magnetic field strength is correlated with density.

\citet{heiles2012} concluded that the Warm Ionized Medium, and possibly the Warm Partially Ionized Medium are responsible for most of the Faraday rotation in the ISM. In this paper, we examine the magnetic field in a nearby structure of ionized intermediate-velocity (IV) gas. \citet{wakker2004} defines IV gas observationally as gas with a velocity with respect to the Local Standard of Rest (LSR) that is between $35$ and $90\ \kms$ outside the velocity range expected for Galactic rotation. In this paper we include in the term IV gas any part of the ISM along the line of sight that moves with a velocity more than $35\ \kms$ with respect to its immediate surroundings.

The association of Faraday rotation with IV gas is significant because it is a kinematically defined component of the ISM that represents a modest fraction of the mass of both neutral and ionized phases of the ISM. Interstellar gas that has recently experienced acceleration, and possibly compression, by the passage of a shock, becomes IV gas until it is decelerated through interaction with its surroundings. Intermediate-velocity gas with this origin may have a higher density and stronger magnetic field, resulting in enhanced Faraday rotation. Whether or not this particular origin applies to the IV gas analyzed here is beyond the scope of this paper.

\vfill 

\section{An IV filament in the local ISM}

\begin{figure*}
\centerline{\resizebox{14cm}{!}{\includegraphics[angle=-90,clip]{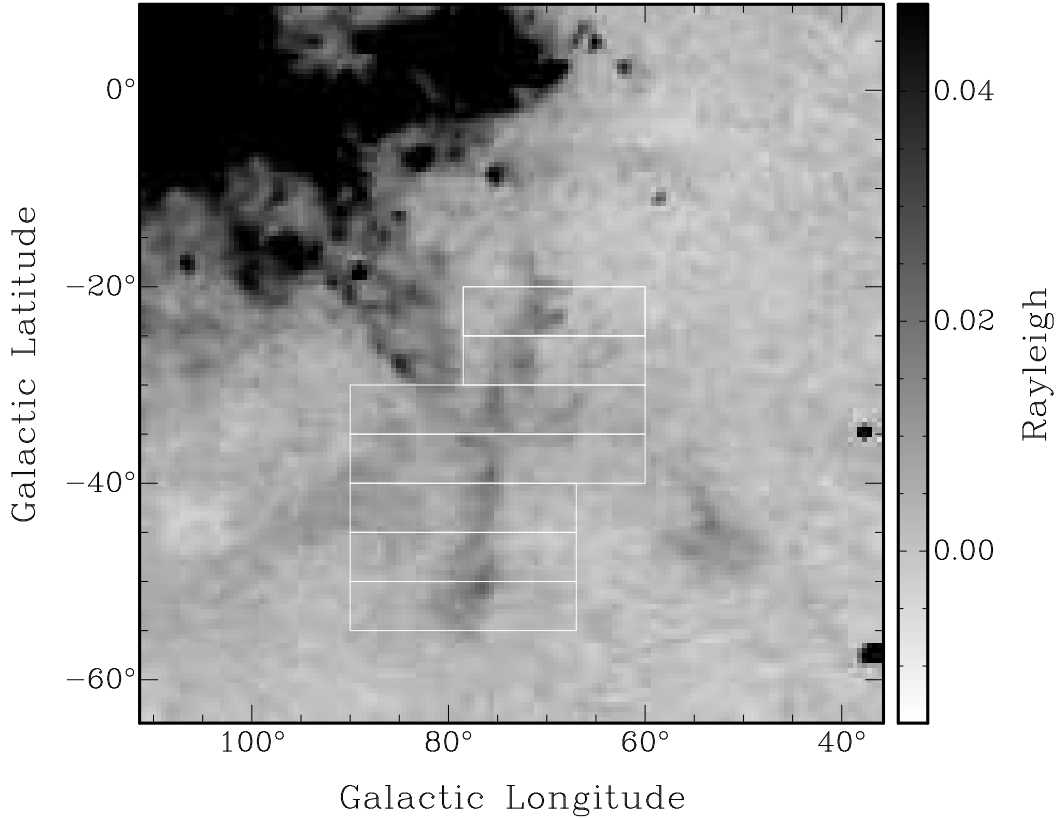}}}
\caption{ H$\alpha$ intensity at $V_{\rm LSR}= -45\ \kms$ from the WHAM survey. The white boxes indicate the regions selected for correlating dispersion measure and rotation measure in Section~\ref{B-sec}. The boxes are numbered 1 - 7 from top to bottom, with boundaries listed in Table~\ref{RMDM-tab}.
\label{filament-fig}}
\end{figure*}

\begin{figure*}
\centerline{\resizebox{14cm}{!}{\includegraphics[angle=-90,clip]{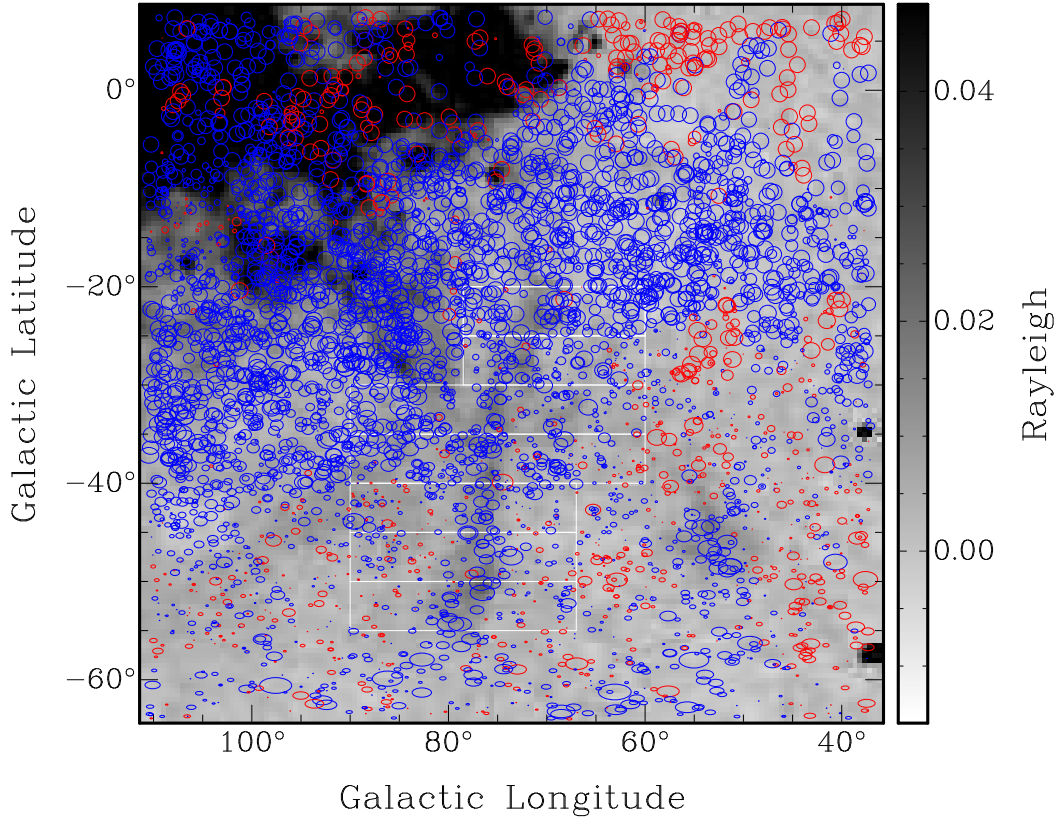}}}
\caption{ Same as Figure~\ref{filament-fig}, with rotation measures from \citet{taylor2009} indicated as circles. Red circles indicate positive or zero $RM$, blue circles indicate negative RM. The size of the circle is proportional to $|RM|$, up to a maximum of $70\ \radm$. The scaling of the circles is chosen to visualize excess negative $RM$ associated with the H$\alpha$ emission at $V_{\rm LSR}= -45\ \kms$. Note the excess Faraday rotation at $(l,b)=(53\fdg5,-46\fdg5)$, associated with a cloud with the same velocity as the filament.
\label{filament_RM-fig}}
\end{figure*}

Figure~\ref{filament-fig} shows an image of H$\alpha$ emission
at $V_{\rm LSR} = -45\ \kms$ from the WHAM survey \citep{haffner2003}
in the direction $l\approx 75\degr$ South of the Galactic plane.  A
long vertical filament that is also visible in the velocity-integrated
sky maps of \citet{hill2008} extends from $(l,b)=(70\degr,-22\degr)$ to
$(l,b)=(78\degr,-55\degr)$ \citep{ade2015}. A detached cloud with the
same velocity as the filament is centered at $(l,b)=(53\fdg5,-46\fdg5)$. 
This cloud was not mentioned by \citet{ade2015}, but it
is almost certainly related because it has the same velocity. 
The region between this cloud and the
filament contains faint H$\alpha$ emission, intensities $\lesssim
0.008$ R per velocity channel, at near the detection limit of the WHAM
survey. The brighter parts of this emission can be seen in
Figures~\ref{filament-fig} and \ref{filament_RM-fig} in boxes 2 and 3,
with fainter emission extending into the low-longitude side of box 4.

\begin{figure}
\centerline{\resizebox{10cm}{!}{\includegraphics[angle=0]{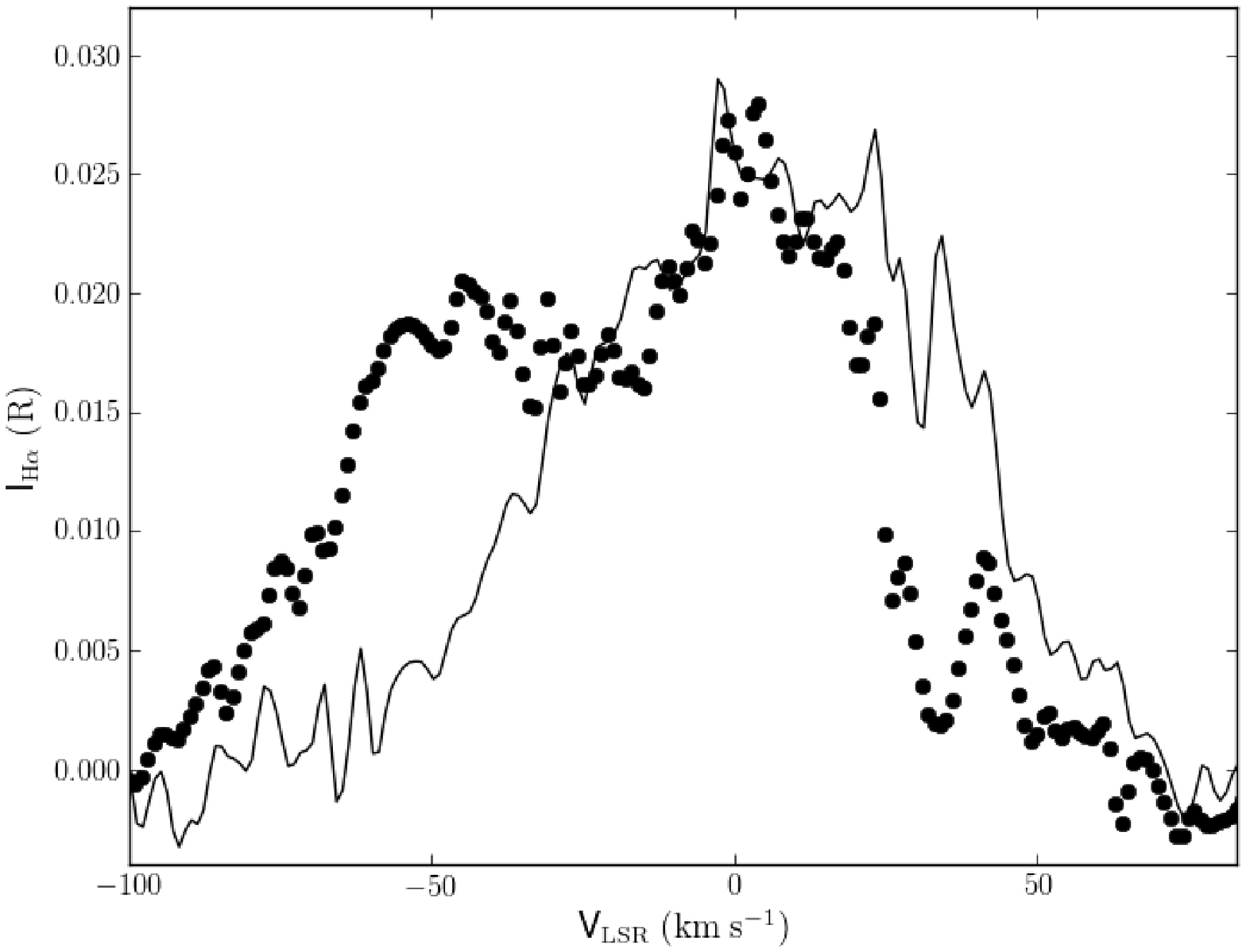}}}
\centerline{\resizebox{10cm}{!}{\includegraphics[angle=0]{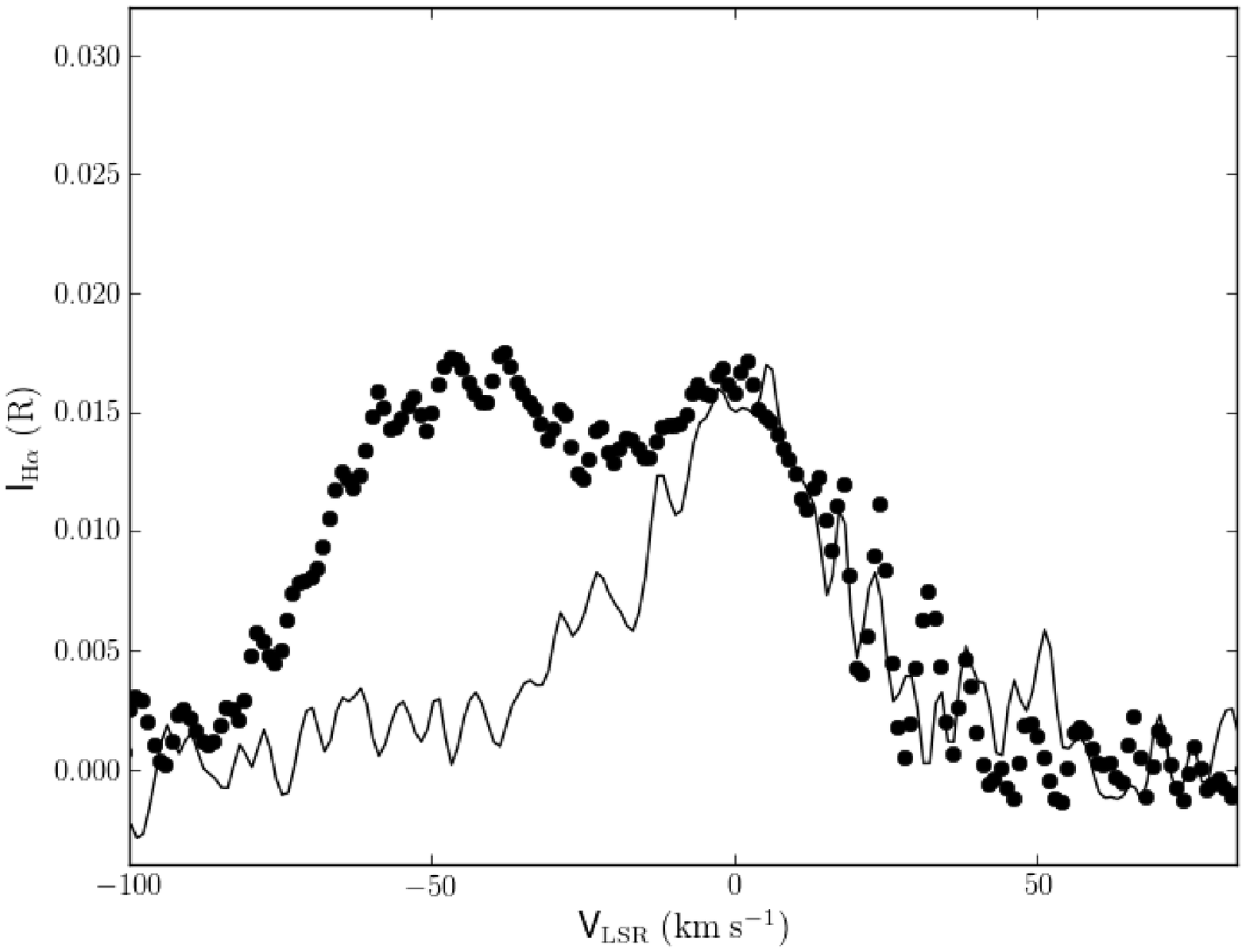}}}
\caption{ WHAM H$\alpha$ spectra at $(73\fdg0,-28\fdg0)$ (top) and $(76\fdg0,-43\fdg5)$ (bottom). Dots represent the spectrum at a position on the filament. Solid lines represent spectra taken at positions with $3\degr$ smaller longitude. The spectral resolution is $12\ \kms$. Emission from the filament is centered around $V_{\rm LSR} = -45\ \kms$. 
\label{filament_profile-fig}}
\end{figure}

\begin{figure*}
\centerline{\resizebox{14cm}{!}{\includegraphics[angle=-90]{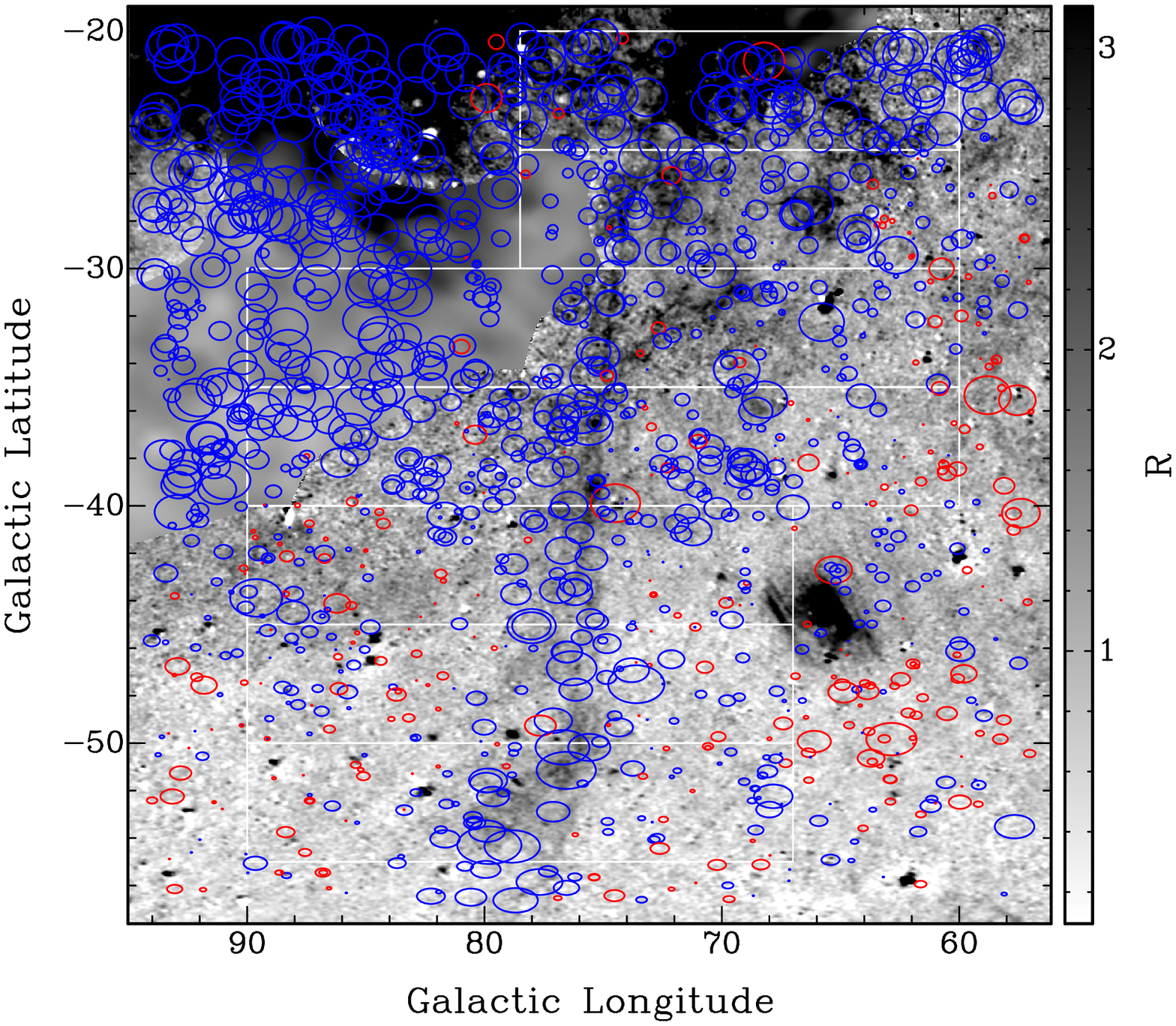}}}
\caption{ Total H$\alpha$ intensity from the composite image from \citet{finkbeiner2003}. The high-resolution portions of this image were provided by the VTSS survey \citep{dennison1998} with angular resolution $\sim 2\arcmin$. 
 The white boxes are the same as in Figure~\ref{filament_RM-fig}.
\label{filament_RM_HR-fig}}
\end{figure*}

More diffuse emission extends from the filament to higher longitude. This
emission has a somewhat lower velocity around $-30\ \kms$ and it is
more blended with brighter, unrelated, emission centered around the
Galactic plane. Bright emission around $b=0\degr$ in
Figure~\ref{filament-fig} is the wing of the bright main disk
component with peak brightness $\sim 3\ \rm R$ per velocity channel,
and center velocity between $0$ to $20\ \kms$.

\citet{ade2015} reported polarized synchrotron emission adjacent to
the H$\alpha$ emission, but not co-located with it. These authors
noted that the free-free opacity of the filament at 1.4 GHz is likely
far too small to explain this minimum in synchrotron intensity by
absorption. We do not investigate the polarized diffuse synchrotron
emission here, but we will address this in a future paper with data
from the GALFACTS survey \citep[at 1.2-1.5 GHz][]{taylor2010}, which will allow a
detailed analysis of the diffuse polarized synchrotron emission.

Figure~\ref{filament_profile-fig} shows two sample spectra at
positions coinciding with the filament, as well as spectra at nearby
offset positions. The spectra show two major velocity components
that are well-resolved at the $12\ \rm \kms$ velocity resolution of
the WHAM survey. We estimate that the typical error in the
velocity-integrated emission measure of the filament derived from the
difference of an ``on'' spectrum with and ``off'' spectrum is
approximately 10\% to 20\%. While the profile of the filament has a
fairly constant peak brightness around $0.017\ \rm R$ per velocity
channel, the brightness of the velocity component centered near
$V_{\rm LSR} = 0\ \kms$ decreases rapidly with distance from the Galactic plane. The
contrast of the filament with its surroundings is stronger farther
from the Galactic equator.

The large extent of the filament in Galactic latitude suggests it is a
nearby structure. If the negative velocity is interpreted as Galactic
rotation, assuming a flat rotation curve with $R_0 = 8.5\ \rm kpc$ and
$V_0 = 220\ \kms$, the kinematic distance is 8 kpc. We reject
  this interpretation because of the lack of related emission at lower
  latitude. The scale height of $\sim 1$ kpc for the WIM corresponds with Galactic latitude
  $7\degr$ at a distance of 8 kpc. The extinction maps of
  \citet{schlegel1998} indicate a mean colour excess $E_{\rm B-V} =
  0.72$ at $b=-5\degr$ and $E_{\rm B-V} = 0.25$ at
  $b=-10\degr$ in the longitude range $70\degr< l < 90\degr$. These
  reddening values correspond to $A_{\rm R}=1.82$ and $A_{\rm R}=0.63$
  respectively. Assuming the brightness of the main disk emission is
  generally higher closer to the Galactic plane, the lack of detectable emission
  at these latitudes at $V_{\rm LSR}=-45\ \kms$ cannot be attributed to interstellar extinction,
  but it is consistent with Galactic rotation.

We suggest the filament is either associated with an extended complex
of neutral IV gas detectable in the 21-cm line, the Perseus-Pisces IV
arch \citep{wakker2004} that has a similar velocity, or that it is
associated with a set of cavities in the local ISM at
the periphery of the Local Bubble, detected in extinction measurements
of stars in the solar neighborhood by \citet{lallement2014}. The
latter implies a distance of 200 to 400 pc.  The distance of the
Perseus-Pisces IV cloud is not very well constrained. \citet{wakker2001} lists an upper limit to the distance of 1.1 kpc
  ($|z|< 0.9\ \rm kpc$) based on a single line of sight towards the star HD
  215733 at $(l,b) = (85\fdg16,-36\fdg35)$, which is inside the area
  analyzed in this paper. We adopt the distance range 200 pc to 400
pc in this paper. The angular thickness of $3\degr$ (FWHM) corresponds
to 10 to 20 pc at a distance of 200 and 400 pc respectively, and the
length of $35\degr$ corresponds with a projected length of 120 to 240
pc respectively.

If the filament is associated with the Perseus-Pisces IV arch, it is
interesting to note that this particular part of the larger IV complex
is mostly ionized. There are a few distinct HI clouds with column
density of a few $10^{19}\ \rm cm^{-2}$ in the Leiden-Dwingeloo survey
\citep{hartmann1997} and the GALFA survey \citep{peek2011} with the
same velocity that coincide with parts of the filament with peak
brightness temperature in the range 1 to 2 K.  The brightest HI
co-located with the filament is found in clouds at $(l,b) =
(75\fdg5,-33\deg)$ and $(l,b) = (76\fdg5,-42\deg)$. The elongated
morphology and low-surface-brightness emission in the Arecibo data of
\citet{peek2011} suggests this HI is associated with the filament. It
will be analyzed in detail along with the GALFACTS data in a future
paper.

  Apparent filamentary structure can also arise from a 2-dimensional
  sheet observed on its edge. The presence of other features at the
  same velocity around the filament suggests that the IV gas as a
  whole is not located in an edge-on sheet. We consider it more
  plausible that the IV gas is located in a two-dimensional sheet
  observed at an intermediate angle with the line of sight. This
  explains the correlated line-of-sight velocities for different
  emission across the field. The simplest interpretation is then that
  the filament is a true filament embedded in this
  sheet. Instabilities may cause a sheet to bend in a more edge-on
  orientation locally, but the fact that the line-of-sight velocities
  of the gas are so uniform in two dimensions remains a problem if
  this is the case. We therefore favor the interpretation of this
  structure as a true filament.

Figure~\ref{filament_RM-fig} shows the same region as
Figure~\ref{filament-fig}, with circles indicating rotation measures
from \citet{taylor2009} with typical errors of $10\ \radm$.  The size
of the circles is proportional to $|RM|$ for $|RM|< 70\ \radm$,
saturating for higher $|RM|$.  The filament stands out because of
significant negative $RM$ compared to its surroundings. Towards the
Galactic plane, the contrast between the filament and the background
is less, because of a strong gradient of the background $RM$ in the
direction of the Galactic plane.

Some extended low-surface-brightness H$\alpha$ emission is visible in
Figure~\ref{filament-fig} and Figure~\ref{filament_RM-fig} just South
of the boundary $b\approx -42\degr$. This emission appears more
blended with the wing of bright emission at normal velocities. Its
velocity is approximately $-30\ \rm km\ s^{-1}$, slightly lower than
the central velocity of the filament. Its association with the
filament is not clear.  There is some indication from the rotation
measures that the filament extends beyond the detectable H$\alpha$
emission, South of $b=-55\degr$, and that it may curve back towards
the Galactic plane.  Any related H$\alpha$ emission in this region is
below the surface brightness threshold of the available H$\alpha$
surveys. In this region of the survey, we see artifacts in the
baseline of the WHAM spectrum that limits the accuracy of emission
measures of some structures with a low surface brightness.

Figure~\ref{filament_RM_HR-fig} shows the same $RM$ data as
Figure~\ref{filament_RM-fig} over the composite image from
\citet{finkbeiner2003} that includes images from the VTSS survey
\citep{dennison1998} with the much higher angular resolution of
$1\farcm 6$ but a lower surface brightness sensitivity. The
high-resolution image reveals substructure in the form of straight,
thin filaments. However, the enhanced Faraday rotation is associated
with the extended H$\alpha$ emission with a surface brightness
$\lesssim 1\ \rm R$. The bright HII region at
$(l,b)=(65\fdg5,-44\degr)$ is a background object with velocity
$V_{\rm LSR}=+6\ \rm km\ s^{-1}$. The top
part of Figure~\ref{filament_RM_HR-fig} lacks high-resolution VTSS images
and is filled with WHAM images at a much lower resolution.

In this paper we investigate the magnetic field in this filament and the detached cloud. The
analysis focuses on the WHAM data, because of the higher sensitivity
to low-surface-brightness emission, and the ability to use the
velocity dimension of the data to improve the background subtraction for the
analysis in Section~\ref{vA-sec}.  The errors in individual rotation
measures and background-subtracted emission measures are
substantial. The analysis will therefore involve averages over the
data in the boxes shown in Figures~\ref{filament-fig},
\ref{filament_RM-fig}, and \ref{filament_RM_HR-fig}. 

\section{Results and Analysis}

\subsection{Density-weighted mean magnetic field}

\label{B-sec}

In order to determine the strength of the magnetic field, we require a dispersion measure. 
\citet{berkhuijsen2006} derived a relation between line-of-sight integrated $EM$ and $DM$ for 157 pulsars,
\begin{equation}
DM = 8 (EM)^{1/1.47},
\label{EMDM-eq}
\end{equation}
where $EM = 2.25\, I_{H\alpha,{\rm tot}}$ (assuming a plasma temperature of 8000 K) is the emission measure that
is derived from the extinction-corrected, velocity-integrated
$H\alpha$ intensity, $I_{H\alpha,{\rm tot}}$. In
this sub-section we use the same $EM$ integrated along the line of sight (all WHAM
velocities), and the same extinction correction, in order to derive values for $DM$ in a way that is
consistent with the source of Equation~\ref{EMDM-eq}. Any contribution from foreground or background is subtracted in the subsequent analysis. The values for
$DM$ that we find are in the lower half of the range for the sample
of \citet{berkhuijsen2006}. The scatter around their relation
suggests significant uncertainty, of the order of a factor 3. 

\begin{figure}
\centerline{\resizebox{9cm}{!}{\includegraphics[angle=0]{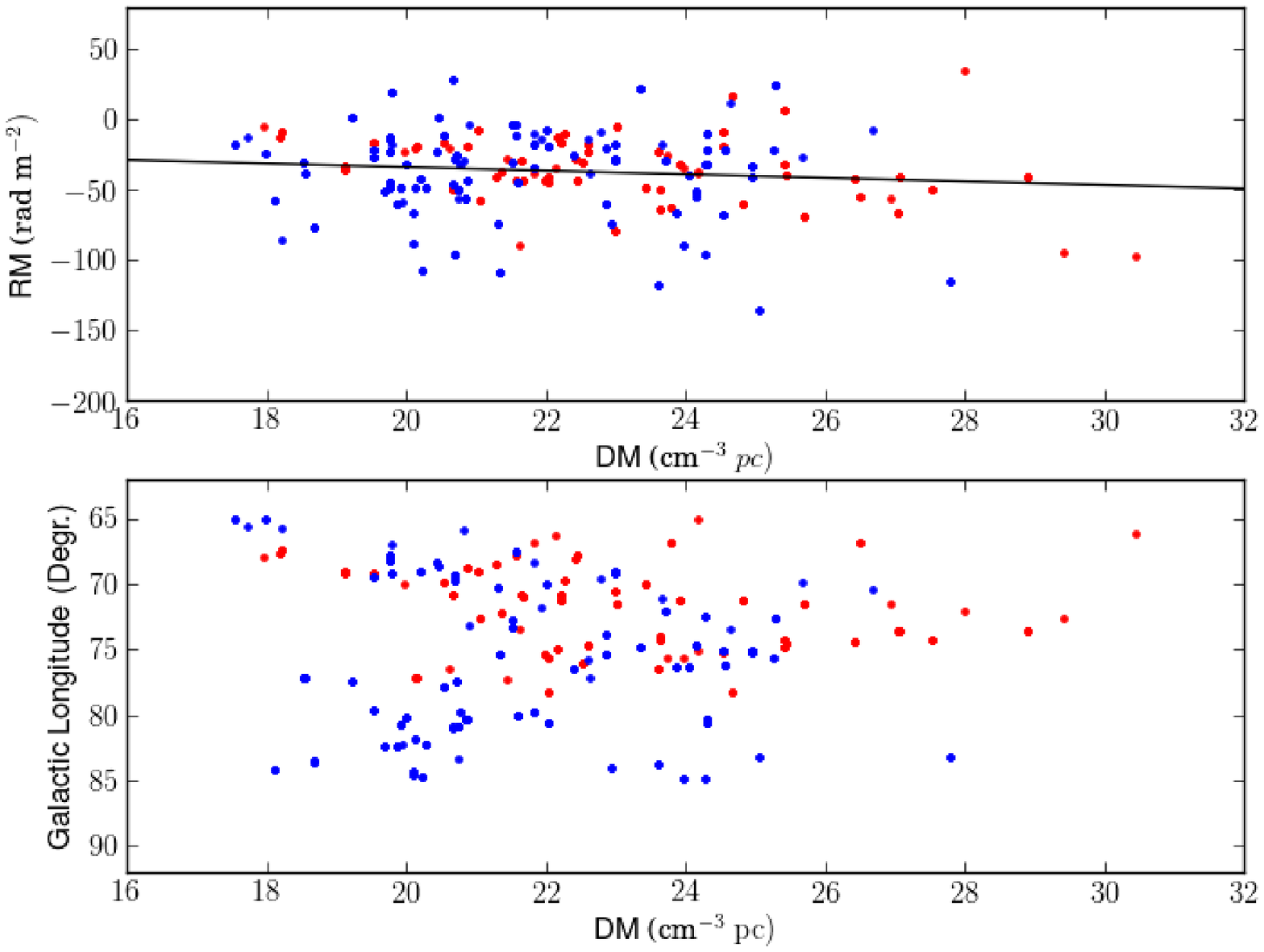}}}
\centerline{\resizebox{9cm}{!}{\includegraphics[angle=0]{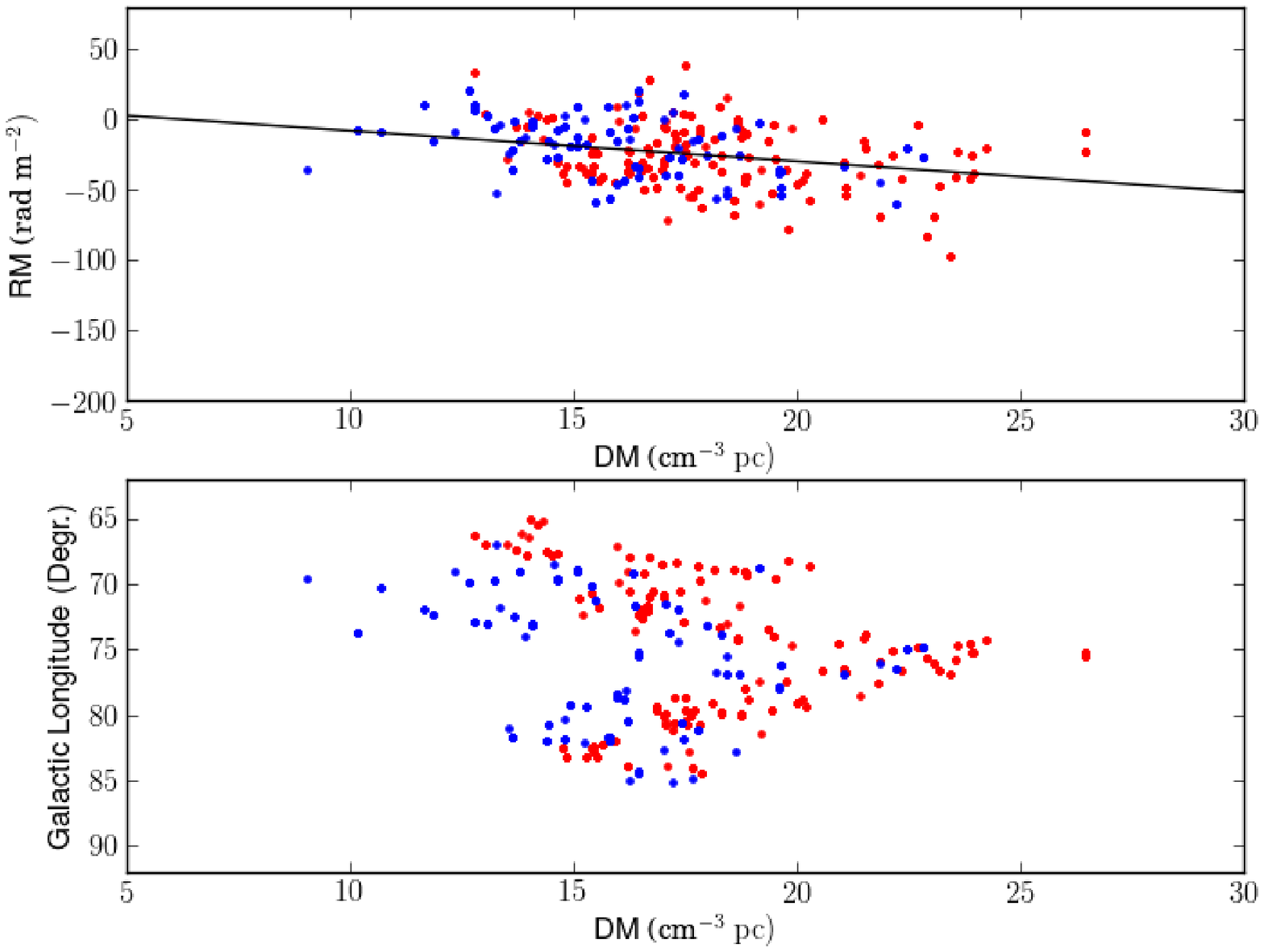}}}
\centerline{\resizebox{9cm}{!}{\includegraphics[angle=0]{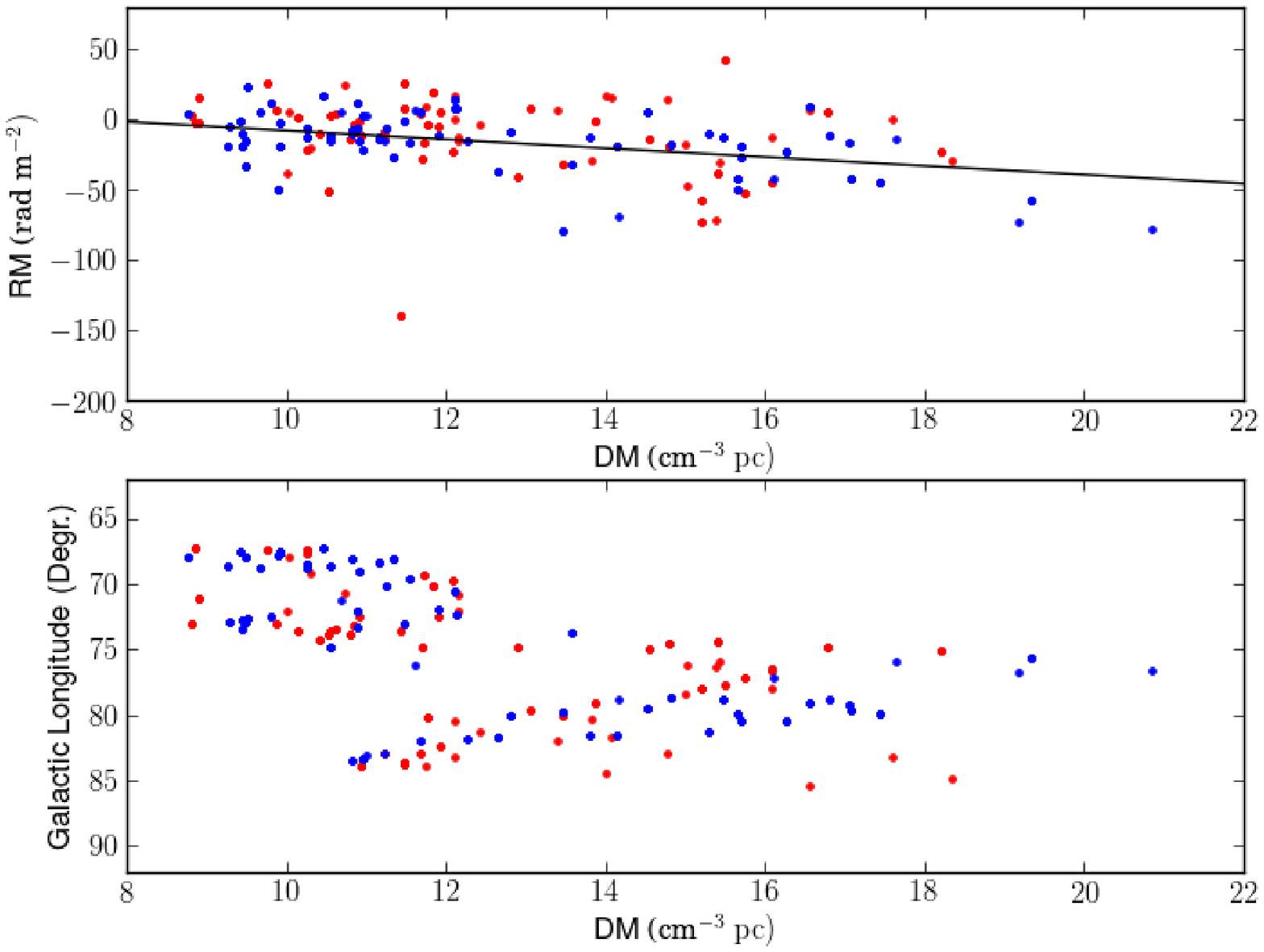}}}
\caption{ Correlation of $RM$ and $DM$, along with $DM$ against Galactic longitude for boxes shown in Figure~\ref{filament-fig}. Top two panels: box 2 (red) and box 3 (blue), central two panels: box 4 (red) and box 5 (blue), and lower two panels box 6 (red) and 7 (blue). The solid lines show the fits to the joint data in each panel (see Table~\ref{RMDM-tab}). The filament is visible as an excess $DM$ near longitude $78\degr$. 
\label{RMDM-fig}}
\end{figure}

Figure~\ref{RMDM-fig} shows examples of the correlation between $RM$
and $DM$, separately for boxes identified in Figure~\ref{filament-fig}. Box 1 was included in our analysis for completeness, but it is not shown in Figure~\ref{RMDM-fig} because the contrast between the filament and its surroundings is small in this box. The scatter in the $RM$-$DM$ relation arises from errors in the individual $RM$ measurements, and from variance in $I_{H\alpha,{\rm tot}}$ that propagates into $DM$. Comparing with the corresponding longitude-$DM$ diagram, we find that a substantial part of the scatter in the $RM$-$DM$ relation is caused by variance in $I_{H\alpha}$. If the scatter in the relation of \citet{berkhuijsen2006} represents true differences between lines of sight, then this scatter propagates in the scale of the $DM$ axis in this figure, and hence a systematic error in the derived magnetic field strength.

In order to derive the mean line-of-sight component of the magnetic
field in the filament, we need to subtract the background $EM$ and the
background $RM$. If we fit a linear relation between $RM$ and $DM$,
the slope of this relation provides the required ratio of the ``on'' -
``off'' quantities. The fits presented here were made to the
  combined data of two adjacent boxes as shown in Figure~\ref{RMDM-fig}. The
resulting magnetic field strengths are listed in Table~\ref{RMDM-tab}.

When fitting the relation between $RM$ and $DM$, we leave the
intercept as a free parameter and express it as the dispersion measure
at $RM = 0$, or $DM_0$, tabulated in Table~\ref{RMDM-tab}. It is
  possible that $DM_0$ is either positive or negative.  A negative
$DM_0$ could indicate that the WHAM images do not include all plasma
along the line of sight, e.g. because of disproportionate extinction
of more distant gas. A positive $DM_0$ can arise if turbulent plasma along the line of sight
adds to the emission measure but not to the mean rotation
measure. We find predominantly positive $DM_0$, because of
  positive $EM$ in the offset positions where the mean $RM$ nearly
  vanishes. 

The density-weighted mean line-of-sight component of the magnetic
field is $2.8\ \mu$G, with a standard deviation of $0.8\ \mu$G. The
standard deviation is less than twice the formal error in the
fits. The magnitude of $B_\|$ is comparable to the strength of
the large-scale Galactic magnetic field. The line of sight is
oriented approximately in the direction of the large-scale Galactic magnetic
field, assuming a pitch angle $\sim 10\degr$.
There is no significant trend of $\langle B_\|\rangle$ with position
  along the filament. This is peculiar, because the angle with the
  line of sight changes by $\sim 35\degr$ over the length of the
  filament, if it is a straight linear structure.  A straight filament
  with a uniform magnetic field oriented along its axis, should show
  systematic variation of $\langle B_\|\rangle$ $\gtrsim 30\%$ or $1\ \rm \mu G$,
  which is not observed. This may indicate that the filament is
slightly curved, or that the magnetic field strength is not uniform,
or that the magnetic field is not aligned with the axis of the
filament.

At this point it is worth noting an interesting combination of 
  observed quantities. First, the projection of the filament's
  axis on the sky is nearly perpendicular to the Galactic
equator. The filament itself may not be perpendicular to the Galactic
plane, but only if its axis makes a significant angle with the line of
sight.

Second, the line-of-sight velocity of the filament is approximately
constant along the filament. This is remarkable considering the
  angular size of the filament. There may be a component of the
velocity perpendicular to the line of sight, but it seems likely that
the velocity vector has a significant component, more than $\sim$ 10
or 20 $\kms$ perpendicular to the axis of the filament, depending on
the true angle between the line of sight and the axis of the
filament. The time it takes for the filament to travel its own
diameter perpendicular to its axis is therefore of the order of a Myr
or less, while the time it takes a sound wave to travel along the
filament is approximately an order of magnitude longer.

Third, the line-of-sight component  of the magnetic field in the filament has the same direction and is similar in magnitude to the large-scale Galactic magnetic field. We do not know whether the mean magnetic field is oriented along the axis of the filament, but
  the thin, straight, filamentary substructure seen in
  Figure~\ref{filament_RM_HR-fig} would be consistent with plasma
  spreading along magnetic field lines oriented along the axis of the
  filament. Enhanced polarized synchrotron emission around the filament
\citep{ade2015} is suggestive of a magnetic field component
perpendicular to the line of sight, at least in the periphery of the
H$\alpha$ filament.  

The combination of these three observations
  raises questions about the dynamical importance of the magnetic
  field in this filament and its interaction with its surroundings. A self-consistent model of this structure should incorporate the complex magnetic field, as well as the peculiar dynamics and morphology. 

In the following section, we use the available data, rotation measure
and emission measure, in a different way to place a lower limit to the
\alfven velocity in the filament.

\begin{deluxetable*}{cccccccc}
\tablecolumns{8}
\tablewidth{0pc} 
\tablecaption{ Mean line-of-sight magnetic field strength }
\tablehead{ Box &  $l$ range$^{a}$ &  $b$ range$^{a}$  &  $\langle B_\|\rangle$ & $DM_0$ & $<RM>$ &  $<EM>$ & $\bar{v}_A$\\     
   &   ($\degr$)          &   ($\degr$)          &  $\rm (\mu G)$ & ($\rm pc\ cm^{-3}$)   & $\rm rad\ m^{-2}$ &  $\rm cm^{-6}\ pc$ & $\rm km\ s^{-1}$}
\startdata
1\&2   &   $60.0$ - $78.5$    &  $-20.0$ - $-30.0$    &  $-3.5 \pm 0.3$   & $9.7\pm 2.6$  & $-14.0 $ & $0.84$ & $16$ \\   
2\&3   &   $60.0$ - $90.0$    &  $-25.0$ - $-35.0$    &  $-3.2 \pm 0.5 $  & $8.8\pm 3.8   $  & $-6.9$ & $0.70$ & $9$ \\   
3\&4   &   $60.0$ - $90.0$    &  $-30.0$ - $-40.0$    &  $-3.4 \pm 0.4$   & $9.1\pm 2.3$  & \ldots & \ldots & \ldots \\   
4\&5   &   $60.0$ - $90.0$    &  $-35.0$ - $-45.0$    &  $-2.5 \pm 0.4$   & $8.0\pm 3.0$ & \ldots & \ldots & \ldots \\   
5\&6   &   $67.0$ - $90.0$    &  $-40.0$ - $-50.0$    &  $-1.3 \pm 0.4$   & $3.6\pm 4.6$  & $-24.0$ & $0.71$ & $32$ \\   
6\&7   &   $67.0$ - $90.0$    &  $-45.0$ - $-55.0$    &  $-3.0 \pm 0.5$   & $8.0\pm 2.5 $  &$-22.0$ & $0.61$ & $33$\\   
\hline
Cloud   &  $42.0$ - $63.0$    &  $-40.5$ - $-51.0$    &  $-6.6 \pm 0.6$   & $12.6\pm 1.8$  & $-32.3$ &$0.51$  & $45$ $^b$\\
\enddata
\tablenotetext{a}{See boxes drawn in Figure~\ref{filament-fig} and Figure~\ref{filament_RM-fig}}
\tablenotetext{b}{Assumes a line-of-sight depth $L = 42\ \rm pc$, for angular diameter $8\degr$ and distance of 300 pc.}
\label{RMDM-tab}
\end{deluxetable*}

\subsection{Density-weighted mean \alfven velocity}

\label{vA-sec}

In view of the considerable uncertainty in some of the steps in the
analysis in Section~\ref{B-sec}, the question rises whether the
available $RM$ and $EM$ can be applied to derive a different quantity of interest
related to the magnetized plasma. We explore the possibility to constrain the \alfven velocity in the region of interest, rather than the magnetic field. As motivation for this approach we can
restate Equation~\ref{phi-eq}, now using SI units for all quantities, in the mathematically equivalent form
\begin{equation}
\phi = 2.63\times 10^{-13} (\mu_e\, m_p)^{1/2} \mu_0^{1/2}\int n_e^{3/2} {B_\| \over \sqrt{\mu_0 \rho} } L dx,
\label{RM-eq-2}
\end{equation}
with $\mu_e\, m_p$ the mean particle mass per free electron, $m_p$ the mass of a proton, $\mu_0 = 4 \pi \times 10^{-7}\ \rm H\ m^{-1}$ the permeability of vacuum, $L$ a length scale that describes the dimension of the object along the line of sight, $x = l/L$ a dimensionless distance along the line of sight, and $\rho = \mu_e m_p n_e$ the mass density of the plasma in kg m$^{-3}$. The reason for introducing the dimensionless distance coordinate $x$ will become clear later. For compact polarized radio sources behind a Faraday rotating screen we can substitute $RM = \phi$ as before. The derivation in this section applies in this case, but not in a case where the polarized radio emission is generated inside the volume of the Faraday rotating plasma.

The \alfven speed in the plasma is defined in SI units as
\begin{equation}
v_A = {B \over {\sqrt{\mu_0 \rho}}},
\label{vA-eq}
\end{equation}
where $B$ is the (total) magnetic field strength. Define the density-weighted mean \alfven speed as
\begin{equation}
\bar{v}_A = {\int n_e^{3/2} v_A dl  \over \int n_e^{3/2} dl}.
\label{vAdef-eq}
\end{equation}
The weighting with electron density $n_e^{3/2}$ is inspired by the integrand of Equation~\ref{RM-eq-2}. If we use Equation~\ref{RM-eq-2} to find an upper bound for $|\phi|$ by replacing $B_\|$ with $|B_\||$ under the integral, and then a further upper bound by replacing $|B_\||$ with $B$, then we obtain the inequality
\begin{equation}
\bar{v}_A \ge \Bigl[{ (\mu_e\, m_p)^{-1/2} \mu_o^{-1/2}  \over 2.63\times 10^{-13}}\Bigr]  {|\phi| \over {\int n_e^{3/2} dl}}
\label{vAlimit-eq}
\end{equation}
The integral in the denominator is not directly available from observations, but we notice its similarity to Equation~\ref{EM-eq}. Define the dimensionless ratio
\begin{equation}
\mathcal{R} \equiv { \int n_e^{3/2} dx \over \Bigl[ \int n_e^{2} dx \Bigr]^{3/4}} = f^{-{1\over 4}} L^{-{1\over 4}} { \int n_e^{3/2} dl \over EM^{3/4}},
\label{R-eq}
\end{equation}
where $L$ is the length scale defined above and $f$ is the filling factor of plasma along the line of sight. The denominator of $\mathcal{R}$ is proportional to $EM^{3/4}$. Use Equation~\ref{R-eq} to eliminate the integral of $n_e^{3/2}$ from Equation~\ref{vAlimit-eq}. The trade-off to eliminate the unknown integral of $n_e^{3/2}$ with the measurable integral of $n_e^2$ is therefore the introduction of a mild dependence on the line-of-sight depth of the object, and the dimensionless factor $\mathcal{R}$ of order unity that can be evaluated for any assumed density structure. Propose a density distribution along the line of sight of the form
\begin{equation}
n_e(x) = n_0 F(x),
\end{equation}
where $n_0$ is a normalizing constant, and F(x) is a dimensionless profile function. The numeric value of $\mathcal{R}$ can be evaluated for any proposed density profile. For a uniform density we find $\mathcal{R}=1$, and for a Gaussian density profile we have $\mathcal{R}=1.0755$, if $L$ is the FWHM diameter of the density distribution. The difference between a constant density and Gaussian density distribution introduces uncertainty of $<10\%$ in $\mathcal{R}$. Expressed in practical units, we find a lower limit for the density-weighted mean \alfven velocity
\begin{eqnarray}
\bar{v}_A \ge ( 0.820\ {\rm km\ s^{-1}})  \mu_e^{-{1/2}} f^{-{1/4}} \nonumber \times \\ \times  \Bigl({L \over 100\ {\rm pc}}\Bigr)^{-{1/4}} \mathcal{R}^{-1} \Bigl[  {|\phi| \over EM^{3/4}} \Bigr]
\label{mean_vA-eq}
\end{eqnarray}
The quantities $RM$ and $EM$ are directly observed. For the other factors, some assumptions need to be made. We discuss each factor in Equation~\ref{mean_vA-eq} in sequence for the filament, to illustrate the conditions under which Equation~\ref{mean_vA-eq} may provide an interesting lower limit. The small exponents of some of these factors mitigate the impact of the assumptions on the result. 

The mean molecular weight per free electron, $\mu_e$, depends on ionization fraction and chemical abundance of the plasma. Assuming hydrogen is 90\% ionized \citep[the lower limit for the WIM quoted by][]{haffner2009}, and the plasma has approximately solar abundances, we apply $\mu_e \approx 1.5$. If hydrogen is fully ionized, $\mu_e = 1.4$, so the uncertainty in the ionization fraction is insignificant compared to other factors in the lower limit to $\bar{v}_A$.

The filling factor $f$ of the plasma enters Equation~\ref{mean_vA-eq} in conjunction with the line-of-sight depth $L$ through the introduction of the factor $\mathcal{R}$. The global mean filling factor $\gtrsim 0.2$ \citep{haffner2009} for the WIM is an average over distances of a few kpc. It is consistent with this global mean filling factor of the WIM to assume a higher filling factor, as high as $f=1$, for the filament. A filling factor $f < 1$ assumes that part of the volume of the filament is occupied by predominantly neutral gas. Although some faint $21$-cm line emission is observed, we adopt $f=1$ here. A significantly lower value for $f$ could raise the lower limit to $\bar{v}_A$. 

Some constraint on the filling factor comes from the lack of variance of $RM$ added by the filament. The effective beam size for the $RM$ measurements is the true angular size of the radio sources ($\lesssim 20\arcsec$). If the filling factor of the plasma is very small, the $RM$ measurements would hit or miss regions with plasma, thus increasing the dispersion of $RM$. Instead, we find that $RM$ is correlated with H$\alpha$ intensity averaged over the $1\degr$ beam of the WHAM survey, and that the enhancement of $RM$ coincides with the diffuse emission in Figure~\ref{filament_RM_HR-fig}. These observations are consistent with a high filling factor of the plasma in the filament.

The value of $\mathcal{R}$ is not a major source of uncertainty compared with other factors in Equation~\ref{mean_vA-eq}. Its value can be constrained in the future with sensitive high-resolution H$\alpha$ images. Figure~\ref{filament_RM_HR-fig} does not show a clear central increase of the surface brightness. For the present, we adopt $\mathcal{R}=1$.

Finally, the value of $L$ requires an assumption about the 3-dimensional shape of the object. As explained before, we favor the geometry of a filament. The FWHM cross section in the H$\alpha$ image is $3^\circ$, which corresponds to a diameter $D = (5\ \rm pc)\, (d/100\ \rm pc)$. If the axis of the filament makes an angle $\theta$ with the line of sight, the path length $L$ through the filament is $L = D/\sin \theta$. The median value for a random distribution is $\theta = 60\degr$. Assuming the nominal distance $d=300\ \rm pc$, we find $L = 18\ \rm pc$ with an uncertainty of $\sim 50$\% because of the uncertainty in the distance that translates into an uncertainty of order 20\% in the limit to $\bar{v}_A$. If we assume that the line of sight is more perpendicular to the axis of the filament, or if we assume a smaller distance, the lower limit to $\bar{v}_A$ will be higher than stated. 

Lower limits to $\bar{v}_A$ are listed in Table~\ref{RMDM-tab}. Box 4 was excluded from this analysis because of the patchy $RM$ structure in this box. In this case we applied no extinction correction to the H$\alpha$ emission, because here $EM$ refers only to the nearby filament that should experience very little extinction. These lower limits to $v_A$ may be compared with the sound speed in a gas with mean molecular weight $\mu$ and temperature $T$,
\begin{equation}
c_S = \Bigl({k T \over \mu m_{\rm H} } \Bigr)^{1/2} = (8.12\ {\rm km\ s^{-1}})\ \mu^{-{1/2}} \Bigl({T \over 8000\ \rm K}\Bigr)^{1/2}.
\end{equation}
The lower limits to $\bar{v}_A$ therefore suggest that the \alfven velocity is similar to or larger than the sound speed $c_S \approx 10\ \kms$ in the filament. The same is often expressed in terms of the plasma $\beta$, the ratio of the thermal pressure $p_{\rm t}$ to the magnetic pressure $p_{\rm B}$,
\begin{equation}
\beta = {p_{\rm t} \over p_{\rm B}} = {c_S^2 \over v_A^2}.
\end{equation}
The lower limit to $v_{A}$ from Equation~\ref{mean_vA-eq}
implies $\beta \lesssim 1$ in the filament, and even $\beta
\lesssim 0.1$ in the boxes farthest from the Galactic plane, where the
contrast in $RM$ between the filament and its surroundings is
higher. Values $\beta \lesssim 1$ indicate that magnetic pressure
dominates over gas pressure. The magnetic field will then restrict
motion of the plasma perpendicular to the field lines. This favors
the formation of filamentary structure aligned with the magnetic
field, as seen in Figure~\ref{filament_RM_HR-fig}.

\section{Discussion}

\subsection{A lower limit to the \alfven speed}

The previous sections derived the density-weighted mean line-of-sight component of the magnetic field, and a lower limit to the density-weighted mean \alfven speed in an ionized filament of IV gas in the local ISM. We propose the latter as an alternative quantity that can be derived from the available data instead of the mean magnetic field strength. If one used the mean magnetic field strength derived in Section~\ref{B-sec} to derive a lower limit to the \alfven velocity, deriving the density indirectly from $EM$ and an assumption about $L$, one would find
\begin{eqnarray}
\bar{v}_A \ge (1.4\ {\rm km\ s^{-1}}) \Bigl( {B_{\|} \over 1 {\rm \mu G}} \Bigr) \Bigl({n_e \over 1\ {\rm cm^{-3}}}\Bigr)^{-{1/2}} \nonumber \\ \sim {(RM) L^{1/2} f^{1/2} \over (DM)^{3/2}}.
\label{vA_from_Bpar-eq}
\end{eqnarray}
Substituting values from Table~\ref{RMDM-tab} and the preceding section leads to a lower limit $\bar{v}_A > 6\ \kms$. Equation~\ref{vA_from_Bpar-eq} also refers to a density-weighted mean \alfven velocity, but with a different density weighting Equation~\ref{vAlimit-eq}. Equation~\ref{vA_from_Bpar-eq} is also more sensitive to assumptions about the line-of-sight depth and in particular the conversion from emission measure to dispersion measure. An implicit assumption in Equation~\ref{vA_from_Bpar-eq} is that the relation of \citet{berkhuijsen2006} applies equally to the filament as it does to the kpc scale lines of sight for which it was derived. This may be questionable when the object of interest is an individual WIM cloud, with its specific physical conditions. Another potential source of error is that the mean extinction correction applied for the complete line of sight is too large for the nearby filament.

If magnetic field strength is correlated with density, systematic errors in the value of $B_\|$ can occur \citep{beck2003}. The averaged quantity $B n_e^{-1/2}$ will be less correlated with density than $B$ itself if $B$ and $n_e$ are positively correlated. Equation~\ref{mean_vA-eq} is also less sensitive to the filling factor of the Faraday rotating plasma than Equation~\ref{vA_from_Bpar-eq}.

Equation~\ref{mean_vA-eq} assumes a constant filling factor and ionization fraction along the line of sight. The assumption of a constant ionization fraction is made when $\mu_e$ is pulled out of the integral in Equation~\ref{RM-eq-2}. This means that Equation~\ref{mean_vA-eq} should not be applied in situations where the ionized fraction varies significantly, for example over long line of sight distances through the ISM, or in other situations where different phases of the ISM are suspected to contribute to the total Faraday rotation. The present analysis focuses on a small-scale structure in the WIM, for which it is easier to justify a large filling factor of ionized gas and a constant ionization fraction. In applications where the filling factor or the ionization fraction is small, the lower limit derived from Equation~\ref{mean_vA-eq} will be more uncertain accordingly.

With the preceding caveats in mind, we briefly consider other applications of Equation~\ref{vAlimit-eq}. Regions A and C in \citet{simard1980} are extended structures identified in all-sky maps of Faraday rotation that are not associated with observable emission in excess of the smooth background emission. They appear at Galactic latitude $|b| \sim 20^\circ$ as extended regions with a well-defined mean $|RM|\sim 25\ \rm \radm$ and emission measure $EM \sim 3\ \rm cm^{-6}\ pc$. Without more information, we treat these regions as completely ionized. Equation~\ref{mean_vA-eq} then results in a lower limit of $\bar{v}_A \gtrsim 3\ \rm km\ s^{-1}$ for line-of-sight depths in the range 1 to 5 kpc. This limit is considerably weaker than the limits derived in Table~\ref{RMDM-tab}. The main reason for this is not the long line of sight, but the ratio $RM/EM^{3/4}$, which is much higher for the filament after subtraction of the background. We can therefore expect the limit to the \alfven speed to be most useful for small-scale structures in the local WIM at high latitude, that can be identified with $RM$ structure despite its small emission measure. Future surveys of Faraday rotation will increase the density of $RM$ measurements. When combined with high-resolution, high-sensitivity H$\alpha$ imaging, these surveys can provide more insight in the dynamical importance of magnetic fields in the ISM on scales of tens of pc.

A completely different application of Equation~\ref{mean_vA-eq} is Faraday rotation of background sources by galaxy clusters. In this case, $EM$ is derived from X-ray observations of thermal bremsstrahlung of the intra-cluster gas. We illustrate this with a numeric example. For $RM$ in the range of 100 to 200 $\radm$ and impact parameter 0.5 Mpc from the cluster center \citep{clarke2001}, electron density $n_e \sim 3\times 10^{-4}\ \rm cm^{-3}$ \citep{croston2008}, $L = 1\ \rm Mpc$, we expect $RM$ and $EM$ to lead to lower limits $\bar{v}_A \gtrsim 50 - 100\ \kms$. These lower limits are well below the sound speed in a $10^7$ K plasma. However, if the observed $RM \sim 100\ \radm$ is the sum of turbulent cells with a scale of 10 kpc, \citep[the value adopted by][]{clarke2001}, then the number of cells along the line of sight would be of order 100. The typical $RM$ per cell would be approximately $10\ \radm$, for an $EM$ of $9 \times 10^{-4}\ \rm cm^{-6}\ pc$ per cell (assuming again $n_e = 3\times 10^{-4}\ \rm cm^{-3}$). Applying Equation~\ref{mean_vA-eq} to a single turbulent cell ($L=10^4\ \rm pc$) would lead to $\bar{v}_A \ge 4.2 \times 10^2\ \kms$, which is not much below the sound speed in a plasma with temperature several $10^7$ K. These numbers serve to illustrate that the limit to the density-weighted mean \alfven velocity introduced here has the potential to be a used to constrain equipartition magnetic field strength in clusters of galaxies, although this estimate will depend somewhat on the assumed turbulent cell size.

\subsection{Faraday rotation by IV gas}

The results presented here suggest that IV gas may be a significant
locus for Faraday rotation in the ISM. \citet{heiles2012}
reviewed possible contributions to Faraday rotation by different
phases of the ISM, but these authors did not discuss IV gas,
which is kinematically different but not related to one specific
  phase of the ISM. Considering that other large-scale $RM$
structures at high latitude are associated with the IV Arch and the
wall of the Orion-Eridanus superbubble \citep{stil2011}, a possible
correlation between $RM$ structure and IV gas
should be considered. The importance of magnetic fields in the
dynamics of this gas and the way it interacts with its surroundings,
is of interest for dynamo theories applied to disks of spiral galaxies
\citep[e.g.][]{parker1971}. Association of Faraday rotation with
such a specific part of the ISM would also be significant for models
of the propagation of diffuse polarized synchrotron emission in
galaxies \citep[e.g.][]{sokoloff1998}.

IV gas is readily identified when its line-of-sight
velocity is outside the range predicted by Galactic rotation. Even at
high Galactic latitude, it is not practical to get a complete sample
of IV gas, if only because of the necessary
projection of the velocity along the line of sight. A general
correlation between IV gas along the line of sight
and $RM$ is therefore not easily established experimentally.

We may ask whether it is plausible that a significant part of $RM$
structure on small angular scales can be produced by structures
similar to the filament discussed here. Over larger distances,
  such structures would contribute to the standard deviation of $RM$
  of extragalactic sources. From the $RM$ structure function images
of \citet{stil2011} we find that the variance of $RM$ on angular
scales of $10\degr$ at latitude $b \approx 50\degr$ is approximately
twice the $RM$ variance on angular scales of $1\degr$ at latitude $b
\approx 15\degr$. Expressed in terms of the standard deviation of $RM$
this represents an increase from $\sigma_{\rm RM} \approx 16\ \radm$
at $b = 50\degr$ to $\sigma_{\rm RM} \approx 22\ \radm$ at
$b=15\degr$. The diffuse H$\alpha$ intensity increases from 0.5 - 1 R
at $b=50\degr$ to 3 - 5 R at $b=15\degr$, while the line of sight
through the Galaxy triples. Assuming a scale height of 1 to 1.5 kpc
for the WIM, the typical line-of-sight distance would increase from
1.3 to 2 kpc at $b=50\degr$ to 4 to 6 kpc at $b=15\degr$.  The angular
scale of $1\degr$ at a distance of 4 to 6 kpc corresponds with a size
of 70 to 100 pc. 

The implied $RM$ and size scale agree well with the estimated size of the
filament discussed here. This also appears consistent with the
  conclusion of \citet{stil2011} that the variance of $RM$ on angular
  scales of 10 degrees must originate in the local ISM, while the
  variance on angular scales of 1 degree arises on lines of sight that
  are so long that they reveal the eccentric position of the Solar
  system in the Milky Way.

The above is suggestive but not conclusive evidence that a substantial
fraction of $RM$ variance in the disk-halo interface could be
explained if the typical line of sight at latitude $b=15\degr$
intersects $\sim 2$ IV structures with size
and Faraday depth similar to the filament discussed here.  Denser
sampling of $RM$ from future polarization surveys, in particular
POSSUM with the Australian Square Kilometre array (SKA) Pathfinder
(ASKAP) and a dense RM grid from a future SKA survey will allow
detailed investigation of more distant structures, and correlation
with Galactic longitude and latitude in relation to tracers of star
formation in the Milky Way.

\section{Summary}

The mean line-of-sight component of the magnetic field in a nearby ionized IV filament is found to be $-2.8\ \pm 0.8\ \rm \mu G$. In a nearby associated cloud the mean line-of-sight component of the magnetic field is $-6.6\ \pm 0.6\ \rm \mu G$. There magnetic fields were derived by estimating dispersion measure $DM$ from the observed emission measure $EM$.

In view of the uncertainty in the conversion of $EM$ to $DM$, we propose an alternative use of the available data to derive a lower limit to the \alfven velocity in the plasma, weighted by density $n_e^{3/2}$. The resulting lower limits are in the range $10\ \kms$ to $30\ \kms$, indicating a low plasma $\beta$.
 
Applications and limitations of this new lower limit to the \alfven velocity are discussed. The derived limit provides the strongest constraints for small-scale structures in the local ISM that are completely ionized. It may also be applied to the turbulent gas in galaxy clusters to constrain equipartition magnetic field strengths.

\section*{Acknowledgments}
This research has been made possible by a Discovery Grant from the
Natural Sciences and Engineering council of Canada to Jeroen Stil.
The National Radio Astronomy Observatory is a facility of the National
Science Foundation operated under cooperative agreement by Associated
Universities, Inc. The Virginia Tech Spectral-Line Survey (VTSS), the
Southern H-Alpha Sky Survey Atlas (SHASSA), and the Wisconsin H-Alpha
Mapper (WHAM) are supported by the National Science Foundation.

{}

\end{document}